%% file: article.tex
\documentclass[12pt]{article}
\usepackage{amsmath}
\usepackage{a4}
\usepackage{epsfig}
\usepackage{dcolumn}

\begin{document}

\newcommand{\av}[1]{\left\langle#1\right\rangle}
\newcommand{\ns}{N_S}
\newcommand{\nt}{N_\tau}
\newcommand{\ds}{\Delta S}
\newcommand{\po}{P_0}
\newcommand{\ps}{P_S}
\newcommand{\sqs}{\sqrt {\sigma}}
\newcommand{\pt}{P_\tau}
\newcommand{\tr}{\operatorname Tr}

\setlength{\topmargin}{-12mm}

\def\be{\begin{equation}} 
\def\ee{\end{equation}} 
\def\bea{\begin{eqnarray}}
\def\eea{\end{eqnarray}}

\renewcommand{\thefootnote}{\alph{footnote}}

\begin{flushright}
Version 21-07-08
\end{flushright}

\vspace{2cm}

\begin{center}
{\Large\bf Thermodynamics of $SU(3)$ Gauge Theory in $2 + 1$ Dimensions}
\\[1.8ex]
{\bf P. Bialas$^{1,2}$\footnote{pbialas@th.if.uj.edu.pl},
  L. Daniel$^1$\footnote{daniel@th.if.uj.edu.pl},
  A. Morel$^3$\footnote{andre.morel@cea.fr}, 
B. Petersson$^{4,5}$\footnote{bengt@physik.uni-bielefeld.de}} \\[1.2mm]
$^1$ Inst. of Physics, Jagiellonian University\\
ul. Reymonta 4, 30-059 Krakow, Poland \\[1.2mm]
$^2$ Mark Kac Complex Systems Research Centre\\
Jagiellonian University, Reymonta 4, 30--059 Krakow, Poland\\[1.2mm]
$^3$Institut de Physique Th\'eorique de Saclay, CE-Saclay \\
F-91191 Gif-sur-Yvette Cedex, France\\[1.2mm] 
$^4$ Fakult\"at f\"ur Physik, Universit\"at Bielefeld \\
P.O.Box 10 01 31, D-33501 Bielefeld, Germany \\[1.5ex]
$^5$ Humboldt-Universit\"at zu Berlin, Institut f\"ur Physik, \\
Newtonstr. 15, D-12489 Berlin, Germany
\end{center}

\vspace{1.5cm}

\begin{abstract}
  The pressure, and the energy and entropy densities
are determined  
  for the $SU(3)$ gauge theory in $2 + 1$ dimensions from lattice Monte
  Carlo calculations in the interval $0.6\leq T/T_c \leq 15$.
  The finite temperature lattices simulated have temporal extent
  $\nt = 2, 4, 6$ and 8, and spatial volumes $\ns^2$ such that
  the aspect ratio is $\ns/\nt =8$. To obtain the thermodynamical quantities, we calculate the averages of
  the temporal plaquettes $\pt$ and the spatial plaquettes $\ps$ on these lattices.
  We also need the zero temperature averages of the plaquettes $\po$, 
  calculated on symmetric
  lattices with $\nt=\ns$. We discuss in detail the finite size
  ($\ns$-dependent) effects. These disappear exponentially. For the zero temperature lattices we find that
  the coefficient of $\ns$ in the exponent is of the order of the glueball mass. On the finite temperature
  lattices it lies between the two lowest screening masses.
  For the aspect ratio 
  equal to eight, the systematic errors coming from the finite size effects are much smaller
  than our statistical errors.
  We argue that in the continuum limit, at high enough temperature, the pressure can be parametrized
  by the very simple formula $p=T^3(a-b\,T_c/T)$ where $a$ and $b$ are two constants.
Using the thermodynamical identities
for a large homogeneous system, this parametrization then determines the other thermodynamical
  variables in the same temperature range. 
\end{abstract}

\vspace{4cm}

\section{Introduction}

The determination of the thermodynamical variables in the high
temperature phase of QCD is one of the main goals of lattice gauge
theory. The values of those variables are of particular importance for the
analysis of the data from high energy heavy ion collisions. Furthermore, the
calculation of the free energy by straightforward perturbation theory
is infrared divergent already at three loops. This divergence, which is
connected to the electric screening mass, can be resummed, but at four loops
new infrared divergences connected to the magnetic screening appear,
which cannot be easily resummed. One may use dimensional reduction
\cite{ginsparg,appelquist,reisz,reisz2}, but the calculation
of the free energy in this framework is a formidable task,
which is not yet completely finished \cite{braaten,kajantie,torrero}.
Therefore, at present a non
perturbative theoretical method like lattice gauge theory is the only 
possibility to get quantitative results.
This is particularly necessary below about twice the critical temperature, where 
perturbation theory even improved with dimensional reduction is not expected
to work.

\medskip\noindent The numerical demand of lattice QCD with
fermions makes it difficult to make a continuum extrapolation with
controlled systematical errors. For improved staggered fermions, however,
there exist new data for the timelike lattice spacings $\nt=4,6,8$ 
\cite{cheng,karsch}. In pure SU(3) gauge theory a continuum
extrapolation has been performed a long time ago,
but only up to $T=4T_c$\cite{boyd}.
In order to make contact with the different proposals for resummed
perturbation theory, it would be very interesting to go to higher
temperature. There are some recent results on this
for $\nt=4$ with the Symanzik improved action
\cite{fodor}.

\medskip\noindent In this article we study a closely related
theory, namely pure SU(3) gauge theory in $2+1$ dimensions. 
This theory has many properties in common with $SU(3)$
gauge theory in $3+1$ dimensions. It has linear
confinement and a deconfining phase transition at finite
temperature. The
infrared divergences in perturbation theory are 
stronger so that even the lowest non trivial order for the free energy
is not calculable in straightforward perturbation theory\cite{dhoker}.

\medskip\noindent 
We determine the thermodynamical variables and
the equation of state by
lattice Monte Carlo computations on several lattices.
In our calculations we use the so called integral method
\cite{boyd,engels}. In order to change from the lattice coupling
constant $\beta$ to the temperature $T$, we use a scaling function
derived from values for the string tension $\sigma$ at zero
temperature \cite{legoetal,lego,teper,teper2}. Using the values for the
critical couplings
from Ref.\cite{legoetal},
we may also express the above thermodynamical variables as functions of
$T/T_c$.
The method then gives directly the normalized trace of
the energy momentum tensor, $(\epsilon - 2p)/T^3$
as a function of $T/T_c$, where $\epsilon$ is
the energy density and $p$ the pressure.
Thermodynamical identities in
a sufficiently large and homogeneous system means that the normalized pressure,
$p/T^3$ can be derived directly from the trace through integration
over $T$. From these two quantities further thermodynamical variables,
like the energy density $\epsilon$ and the entropy density $s$ can be
easily calculated. We determine the 
thermodynamical variables in the range $0.6 \leq T/T_c \leq 15$, which
in this theory means that the effective dimensionless coupling
constant $g^2/T$ varies in the interval $ 3\geq g^2/T \geq 0.12$.

\medskip\noindent In Section 2 we present the problem and our adaptation
of the integral method. 
In Section 3 we derive the $\beta$-function
and determine its parameters.
In Section 4 we present the results of our Monte Carlo calculation,
and show that finite spatial lattice size effects are not present in the data. 
In Section 5 we discuss the results.
Section 6, finally, is devoted to
the conclusions.

\section{The method}

In this section we will first recapitulate some well known facts about
thermodynamics and lattice gauge theory in $2+1$ dimensions.
This serves essentially to fix our definitions and notations.   
We then describe the method employed to extract the thermodynamical 
quantities from lattice data, following ref. \cite{boyd,engels}

We start with the Euclidean Lagrangian in the continuum theory,
\begin{equation} 
{\cal L}_E ( {\mathbf A}_\mu (x)) = \frac{1}{2 g^2} tr ({\mathbf F}_{\mu
\nu} (x) {\mathbf F}_{\mu\nu} (x)).
\end{equation}
Here, $x = (x_0, x_1, x_2)$ is a three dimensional Euclidean vector. 
 The dynamical 
variables are the gauge fields ${\mathbf A}_\mu (x)$, which are
hermitean
traceless matrices
belonging to the algebra of the SU(3) group and
\begin{equation} 
{\mathbf F}_{\mu\nu} (x) = \partial_\mu {\mathbf A}_\nu (x) - \partial_\nu
{\mathbf A}_\mu (x) + i [ {\mathbf A}_\mu (x), 
{\mathbf A}_\nu (x) ].
\end{equation}

\noindent
Note that $g^2$ has the dimension of mass in three dimensions.

The thermodynamics of the corresponding quantum theory is 
derived from the partition function, which is formally expressed as
\begin{equation} 
Z (T, V, g^2) = \int {\cal D} {\mathbf A}_\mu (x) e^{-\int^{1/T}_0 dx_0 \,
\int_V d^2 x {\cal L}_E ({\mathbf A}_\mu(x))} ,
\end{equation}
where $T$ is the temperature and $V$ the spatial volume.

From the partition function $Z$ we get the free energy
\begin{equation} 
F (T, V, g^2) = - T \log Z.
\end{equation}
In the following we will use the pressure $p$, and the volume densities of
the free energy, the internal energy and the entropy, and denote those by
$f$, $\epsilon$ and $s$ respectively.
We will assume that we have a large homogeneous system, in which case
\begin{eqnarray}
p (T, g^2) & = & - f (T , g^2), \\
\epsilon (T , g^2 ) & = & T^2 \frac{\partial}{\partial {T}} (p/T), \\
s (T , g^2) & = & \frac{\partial p}{\partial T} = \frac{\epsilon + p}{T}. \label{entropy}
\end{eqnarray} 
It follows that in $2+1$ dimensions we further have for the trace of the energy
momentum tensor $\epsilon - 2p$, 
\be
\frac{\epsilon-2p}{T^3} = T\frac{\partial}{\partial T} \left( \frac{p}{T^3}
\right). \label{trace}
\ee

In fact we can use the left hand side as the basic quantity, from which the other 
thermodynamical quantities can be obtained apart from an integration constant.

In $2\,+\,1$ dimensions and for $g^2 = 0$ (the free gauge theory), the temperature $T$ is the only 
scale in the system, and the dimensionless quantities 
\be
\frac{p}{T^3} , \frac{\epsilon}{T^3} , \frac{s}{T^2},
\ee
are pure numbers.
Thus, in this case the energy momentum tensor is traceless and
\begin{eqnarray}\label{freep}
\epsilon & = & 2 p, \\
s & = & \frac{3p}{T}. 
\end{eqnarray}
The pressure for eight free gluons in two spatial dimensions is easily calculated
to be 
\be
\frac{p}{T^3} = 8\frac{1}{2\pi}\zeta (3) = 1.5305... \, .
\ee

In this article we consider the theory regularized on a
finite lattice with lattice spacing $a$ and with $\nt$ points in the
(inverse) temperature direction, defined as the 0 direction and $\ns^2$ points in the 
space directions $1,2$. We denote the link variables starting from
the site $x$ in the positive directions by $U_{\mu}(x), \mu=0,1,2 $
and use the standard Wilson
action: 
\begin{eqnarray} 
 S_W (U_\mu (x)) &  = & \sum_P S (U_P) , \\
S(U_P) & = & \beta
\left( 1 - \frac{1}{3} Re \, Tr \, U_P \right), 
\end{eqnarray}
where $P$ denotes one of the $3 \nt  \times \ns ^2$ plaquettes on the
lattice and $U_P$
is the product
of the $U$-matrices around the plaquette. 
 One should not confuse the dimensionless coupling constant
$\beta$ with the inverse temperature, which we will not use
with that notation. Defining $U_\mu (x)$ in terms of
$A_\mu (x)$ by \be U_\mu (x) \equiv e^{i \int^{x + a \hat\mu}_x \,
  A_\mu (x^\prime ) d x^ \prime} \cong e^{ia A_\mu (x)} , \ee
one obtains the original Lagrangian from the classical limit of the lattice
Lagrangian,
\begin{eqnarray}
\lim_{a \rightarrow 0}\frac{S(U_P)}{a^4} & =& \frac{\beta}{12}
Tr F_{\mu\nu} (x) F_{\mu\nu} (x), \\
\lim_{a \rightarrow 0}\,a\beta &=& \frac{6}{g^2 }.
\end{eqnarray}
Furthermore we define the temperature and the volume on the lattice by
\begin{eqnarray}
\frac{1}{T} & = & a \nt , \label{T} \\
V & = & (a\ns)^2. \label{V}
\end{eqnarray}

To obtain the thermodynamical quantities we need to calculate the free
energy on the lattice,
\begin{eqnarray} 
 p \, & = & \,-f\,=\,\frac{T}{V} \log Z, \label{pdef} \\
Z & = & \int \prod_{x , \mu} \, d U_\mu (x) e^{-S_W (U_\mu (x))}.
\end{eqnarray}

\noindent
The partition function cannot be calculated directly by Monte Carlo methods.
Here we use the integral approach of Refs. \cite{boyd,engels}. Thus we first calculate
the derivative of $\log Z$ with respect to the lattice coupling constant $\beta$
at fixed $\nt$ and $\ns$. It is given by 
\begin{equation} 
\frac{d\log Z}{d\beta} =  - \nt \ns^2 \av{\ps + 2 \pt} \label {diffp}.
\end{equation}
The negative sign may look strange, but we have not yet subtracted the pressure at zero
temperature.
Note that we define the plaquette average value $\av{P}$ as
\begin{equation} 
\av {P}  = \av{ 1 - \frac{1}{3} Re \, Tr \, U_P}
\end{equation}
and $\ps$ (resp. $\pt$) denotes a plaquette in the $\{1,2\}$
(resp. $\{0,1\}$ or $\{0,2\}$) plane. Averages $\av{\ldots}$ are taken
with respect to the Wilson weight corresponding to given lattice
parameters.  The averages containing either $\pt$ or $\ps$ are always
taken on the finite temperature lattices $\nt<\ns$ and $\av{\po}$ will
always denote average on the zero temperature lattice $\nt=\ns$.

Eqs. (\ref{pdef}, \ref{diffp}) defines $p$ up to a constant with respect to $\beta$,
so that for some $\beta^0$ to be chosen below we find
\begin{equation}
a^3\,p (\beta ,\nt , \ns) = - \int^{\beta}_{\beta^0}
\, d \beta^\prime  \av{\ps+2\pt}_{\beta'} + a^3\,p (\beta^0 ,\nt,\ns).
\end{equation}
We next subtract from this expression its value for $\nt = \ns$, which
for $\ns$ large enough constitutes its zero temperature limit, and thus
consider the quantity
\begin {eqnarray}
&&a^3\, \biggl( p(\beta , \nt , \ns)-p (\beta , \ns , \ns)\biggr)
\nonumber \\
\,&=&\,
\int^{\beta}_{\beta^0}
\, d \beta^\prime  \Biggl(3\,\av{\po}_{\beta'}\,-\,\av{\ps+2\pt}_{\beta'}
\Biggr)  \nonumber \\
&+&a^3\, \biggl( p(\beta^0 ,\nt,\ns)-p (\beta^0 ,\ns,\ns)\biggr).
\end{eqnarray}
The last term in this substracted expression for the pressure is still
unknown, but it can be made negligible by choosing $\beta^0$ in such a
way that $\nt$ is sufficiently larger than the correlation length at
that value of $\beta$. To this extent then $p(\beta^0 ,\nt,\ns$) is
insensitive to the replacement of $\nt$ by $\ns$ and the constant
vanishes.  Under this condition for $\beta^0$, our numerical estimate
of the thermal (i.e. zero temperature subtracted)
 part of the pressure, which we denote by the same symbol $p$, is
\begin{equation}
\frac{p(\beta,\nt,\ns)}{T^3}\,=\,\nt^3\,\int^{\beta}_{\beta^0}
\, d \beta^\prime \ds(\beta',\nt,\ns), \label{platt}
\end{equation}
where  
\begin{equation}\label{eq:ds}
\ds(\beta,\nt,\ns)=3 \av{\po}_{\beta}\,-\,\av{\ps + 2\pt}_{\beta}.
\end{equation}
The above expression for the pressure is evaluated from simulations
performed on $\nt\times \ns^2$ lattices, with the aspect ratio
$\ns/\nt \ge 8$. In section 4 we will show that for this minimum aspect
ratio we are sufficiently close to the thermodynamic limit, i.e.  that
the finite size corrections are smaller than our
statistical errors.

In order to obtain the pressure from (\ref{platt}) as a function of
$T/T_c$, where $T_c$ is the critical temperature of the continuum theory,
 we still need to
relate $T/T_c$ to $\beta$, 
\begin{equation}
 \beta = \beta (T/T_c), \label{betat}
\end{equation}
which will be discussed in Section 3. Finally we choose
$\beta^0=\beta(T/T_c=0.6)$, after checking from the measurements that
$\nt^3$ times the integrand in (\ref{platt}) is negligibly small at
this temperature.  According to (\ref{trace},\ref{platt},\ref{betat}),
the normalized trace of the energy momentum tensor is finally given by
\begin{eqnarray}
\frac{\epsilon - 2 p}{T^3} & = & T \frac{\partial}{ \partial T} 
\left( \frac {p}{T^3} \right)  \nonumber\\ 
 & =&  \nt^3 \,\ds\left(\beta \bigl (\frac{T}{T_c}\bigr ),\nt,\ns \right ) \label{anomaly}
 \,T \frac{d\beta}{dT} .
\end{eqnarray}
From Eqs. \eqref{entropy}, \eqref{platt} and \eqref{anomaly} one
obtains the expressions for the energy and entropy densities
$\epsilon$ and $s$.

\section{ Physical Scales for the Temperature}

We need to express the thermodynamical variables, provided by 
\eqref{platt} and \eqref{anomaly} from lattice simulations, as functions of
the lattice temperature $T$ defined in (\ref{T}). This temperature
has to be measured at some scale related to the continuum theory. The continuum
coupling $g^2$ is such a scale, but it is interesting to consider other
quantities of the same dimension, and thus proportional to $g^2$, 
which have a direct physical interpretation.
Here we will use $\sqs$, the square root of the zero
temperature string tension, and $T_c$, the
critical temperature, both in the continuum theory.

In this section, we construct the corresponding 
scaling functions which, given $\ns$ and $\nt$,
represent the change of variable from $\beta$ to $T/\sqs$, 
$T/T_c$ or $T/g^2$. We start with the former, namely
\be
\beta\,=\,\beta_\sigma(T/\sqs) \quad\text {and its inverse}\quad
\frac{T}{\sqs}\,=\,\beta_\sigma ^{\{-1\}}(\beta). \label{eq:betafn}
\ee
Passing from one scale to the other is just a multiplicative
numerical factor, $T_c/\sqs$, which must be also computed.

In the literature, there exist mainly two sets of data for the quantity
\be
F_{\sigma}(\beta)\,=\,a\,\sqs \label{fsig},
\ee
obtained from lattice simulations in various domains of $\ns$ and
$\beta$. A large set $\beta \in [8:50]$, 
referred to as $A$ in what follows, 
has been covered in \cite{lego, teper}, while recent simulations \cite{teper2}, 
provide high accuracy data at three $\beta$ values only, 14.717, 21 and 40
(set $B$).
Both sets are included in table~\ref{tab2:sigma}. How we use
them is explained below.

\begin{table}
\begin{center} 
\input sigma.tex

\end{center}
\caption{\label{tab2:sigma}
  Data for the string tension.  Each line gives the values of
  $\beta$, $\ns$, and $F_\sigma(\beta)\,=\,a\,\sqs$, and the corresponding reference.} 
\end{table}

For any value of $\beta$ occuring in Table~\ref{tab2:sigma}, using Eqs.
(\ref{fsig},\ref{T}) gives directly
\be
\frac{T}{\sqs}\,=\, \frac{1}{\nt\,F_\sigma(\beta)}.  \label{tsigma}
\ee
However, we need in practice to evaluate
$T/\sqs (\beta)$ for a denser and larger set ($\beta \in [8:400]$) 
of $\beta$ values. This is required in particular to obtain 
the pressure from the integral (\ref{platt}),  in a range of
temperatures  $T/T_c$ from about 0.6 to 15.  This 
we achieve by making fits to the data of Table~\ref{tab2:sigma}.
We use the form
\begin{equation}\label{eq:asigma}
F_\sigma(\beta)  = \frac{c_1}{\beta} + \frac{c_2}{\beta^2} + 
\frac{c_3}{\beta^3}.
\end{equation}
This choice is done for the sole purpose of representing properly
the data, with no theoretical prejudice  about the actual behaviour
of $F_\sigma$ around $1/\beta$=0.
An investigation shows that there exists a systematic although 
small discrepancy between the sets $A$ and $B$, and we found it adequate 
to fit them separately rather than altogether. Replacing the symbol $c_i$
of (\ref{eq:asigma}) by $a_i$ and $b_i$ respectively for $A$ and $B$,
the two fits give

\begin{eqnarray}
 a_1=3.37(1) \quad  a_2=3.90(25)  \quad  a_3=50.1(1.8) \label{afit} \\  
 b_1=3.34(1) \quad  b_2=4.68(50)  \quad  b_3=39.8(5.0) \label{bfit}.
\end{eqnarray}
The relative difference of $1\%$ between $a_1$ and $b_1$ gives the order
of magnitude of the systematic effect due to choosing $A$ or $B$ at
large $\beta$. At low $\beta$, one might worry that the large cubic
term found in both cases casts a doubt on the validity of the
extrapolations. We note however that
table~\ref{tab2:sigma} gives $F_\sigma$ = 0.568(2) 
at $\beta=8.156$ \cite{lego}, less than $3\%$ away from the average
values 0.564 and 0.553 one gets respectively (with larger errors)
from $A$ and $B$. 
Furthermore, as we shall see in the next
sections, such low values of $\beta$ are used only for $\nt\,=\,2$,
which plays a very marginal role in our
main physical results. With that in mind and for
definiteness, we will use the parameters (\ref{bfit}) corresponding to the 
high precision data \cite{teper2}. The numerical value of $T/\sqs$ given
$\beta, \nt$ is thus obtained by
inserting (\ref{eq:asigma}) in (\ref{tsigma}) with $c_i=b_i$,
$b_i$ as in (\ref{bfit}). The errors on $\sqs$ in the input data imply an
error on this change of variable, to be estimated (see below).

Our last task before that is to define $T$ using the scale $T_c$, the critical
temperature in the continuum, that is in the limit $\nt$ and  $\ns \to \infty$.
Since we already know $T/\sqs$  we thus compute $T_c/\sqs$.  
From (\ref{tsigma}), we write
\be
\frac{T_c}{\sqs}\,=\, \lim_{\nt \to \infty}\,
\frac{1}{\nt\,F_\sigma[\beta_c(\nt)]},  \label{tc}
\ee
where $\beta_c(\nt)$ is the critical lattice coupling determined by
lattice simulations at given $\nt$. In this expression, 
the $\ns=\infty$ limit should have been taken first. As will be shown in
section 4 on the basis of our simulations, the value of $\ns$ (8$\times
\nt$) for each $\nt$ is large enough to eliminate any finite $\ns$ 
effect. In order to perform the limit (\ref{tc}), we use the set 
 of $\beta_c$
obtained in \cite{lego} for $\nt$=2, 4 and 6, that is respectively
$\beta_c$=8.155(15), 14.74(5), and 21.34(4)(11). More recent data on
$\beta_c(\nt)$ at $\nt=2,3,4,5$ can be found in
\cite{liddle} and they do not modify our estimate, which is:
\be
\frac{T_c}{\sqs}\,=\,1.00(4) \label{tcsigma}.
\ee
The ratio to this constant of $T/\sqs$  is $T/T_c$, which will be
used as the temperature variable in the rest of the paper.

\medskip\noindent
The error in (\ref{tcsigma}),
as well as any other error quoted in the present and subsequent
sections have been estimated using the so-called {\it bootstrap}
method (see \cite{efron,numrec}). To be concrete, we describe it
in detail for this very typical case. 

The set of initial data, noted $x_j$ for short, consists of values and
errors for $a\,\sqs$ and $\beta_c$ at given values of external variables
$\beta$ or $\nt$. All $x_j$ are assumed to be 
independent and gaussians. We performed the following steps

{\it Step 1}: From the last assumption, generate another equivalent set
of input variables $y_j$.

{\it Step 2}: Fit the coefficients $c_i$ of (\ref{eq:asigma}) to the $y_j$ values.
(Note that in the case where $y_j$ represents the data of \cite{teper2},
the 3 coefficients $a_i$ are exactly determined by the 3 $y_j$. Standard
$\chi^2$ methods would not give any estimate of their errors).

{\it Step 3}: Compute $F_\sigma(\beta)$ 
for $\beta\,=\,\beta_c (\nt)$ 

{\it Step 4}: Fit a quadratic form in $1/\nt$ to the 3 values obtained
for the inverse of $\nt\,F_\sigma(\beta_c (\nt))$. From Eq. (\ref{tcsigma}), 
its value at $1/\nt$=0 is an estimate of $T_c/\sqs$.

{\it Step 5}: Repeat everything from Step 1 many times (100), 
each time with a new set of $y_j$.

Our result (\ref{tcsigma}) gives the value of $T_c/\sqs$ for
$y_i\,=\,x_i$, with an error equal to the standard deviation measured
within the set of estimates obtained at Step 5.

\medskip\noindent
\medskip\noindent
\medskip\noindent

A different way to determine a $\beta$ function of the form
(\ref{eq:betafn}) is to use directly the knowledge of $\beta_c$
available for various $\nt$ values. We expand $\beta$ as a Laurent series in
$a\,g^2$, writing 
\begin{equation} 
\beta\,=\, \frac{6}{a\,g^2}\,+\,b\,+\,c\,a\,g^2\,+\,\cdots
\label{betag2}.
\end{equation} 
The first term corresponds to the classical limit of the lattice
action. We include as many terms as there are known
$\beta_c$ values, and consider the truncated series obtained as
representing the first terms of an asymptotic series. We then replace
the lattice spacing $a$ by $1/(\nt\,T)$ and introduce the scale $T_c$
to rewrite (\ref{betag2}) as
\begin{equation}
 \frac {\beta}{\nt}\,=\,
\frac{6\,T_c}{g^2}\,\frac{T}{T_c}\,+\, \frac{b}{\nt}\,+\,
c\,\frac{g^2}{T_c}\,\frac{T_c}{T}\,\frac{1}{\nt^2}
\,+\,\cdots \label{betatc}.
\end{equation}
This expansion now constitutes an expansion in $1/\nt$, whose 
$1/\nt\,=\,0$ limit gives an estimate of $T_c$ in units of $g^2$
in the continuum limit.

From Eq. (\ref{betatc}) and the values of $\beta_c$ quoted in
\cite{lego} for $\nt=$ 2, 4, 6, namely  8.149(3), 14.74(5), 21.34(11)
respectively, one finds 
\be \frac{T_c}{g^2}\,=\,0.55(2) \,; \quad \quad
b=1.5(6) \,; \quad \quad c\,=\,0.06(50) \label{paramtc}.  
\ee 
We observe that the $1/\nt^2$ term is compatible with zero in the range of
interest, which means that the $\beta$ function  in terms of 
$T/g^2$ is essentially linear. Although the $\beta$ functions
determined via the knowledge of the string tension and via
Eqs. (\ref{betatc}, \ref{paramtc}) need not be the same at finite
$\nt$, we find them very close one to the other, so that they can 
be used equivalently in the present work. 
In practice, an accurate representation of our findings for the
$\beta$-function, valid in the whole range explored in $\beta$ and $\nt$
is
\be
\frac{\beta}{\nt}=3.3\,\frac{T}{T_c}\,+\,\frac{1.5}{\nt}.  \label{fastb}
\ee
This formula is useful in order to jump easily from functions of
$\beta$ to functions of $T/T_c$. Only when precise estimates of 
errors in final results is needed is it
necessary to go through the process described above and apply the
bootstrap method.

\section{Results of the Simulations}
\label{sec:results}

The results of our simulations consist of high precision
plaquette data, that is their average values and statistical errors,  for zero 
and finite temperature lattices. 
These data are shown in the appendix 
Here we present and discuss the quantity 
$\nt^3\,\ds\,=\,\nt^3\,(3\av{P_0}\,-\,\av{2\pt+\ps})$ which was shown in
section 2 to
constitute the basis for all computations of the thermodynamical variables.

\medskip\noindent
An important source of possible systematic errors for the quantities of
physical interest resides in the finite size effects (FSE) at
$\beta$ and $\nt$ fixed due to $\ns$ being finite.  
A reason for this is that although $\av {\tr {U_P}}\,/\,3$ is
of order one, 
very strong cancellations occur in $\nt^3\,\ds$. According to
the  definitions in section 2,    
$\av{P}$ is expected to go to 0 as $\beta \to
\infty$. Although its precise behaviour around $\beta^{-1} \,=\,0$ is
unknown, we assume that it can be numerically represented by a
(probably asymptotic) series in $\beta ^{-1}$. We find that the
linear term is absent and that
$\nt^3\,\ds$ is of the form
\begin{equation} 
 \nt^3\,\ds\,
\approx\, \frac{A}{\beta^2} + O(\beta^{-3}) \label{beta3},
\end{equation}
at large $\beta$, with $A$ a function of $\nt$ and $\ns$.
This behaviour was verified to a high degree of precision from our
data.

We, of course, have to make sure that the finite size effects occuring in the
distinct measurements of $\po,\pt,\ps$ are so small that the
corresponding effects in $\nt^3\,\ds$ lead to acceptable errors,
smaller than the statistical ones.  These finite size effects
should be related to the largest gauge invariant correlation length of
the system, a priori different for $\po$, and for $\ps$ and
$\pt$. In the former case we are in the confined phase at $T=0$,
where the largest correlation length is given by the inverse of the
lowest glueball mass, whose measurement was reported in
\cite{teper}. In the latter case, it is given by the screening length,
measured in \cite{bialas}.

Here we report the results of a systematic study of the FSE observed
in $P_0,\pt,\ps$, from simulations performed, at $\nt$=4 fixed, for a
large set [15:150] of $\beta$ values and $\ns\in [6:48]$.

Each $\av{P}$ at a given $\beta$ was analyzed as a function of $\ns$,
and parametrized as 
\begin{equation} 
\av{P}_{\beta,\ns}\,=\,A(\beta)\,-\,B(\beta)\,\exp(-\mu (\beta)\,\ns)
\label{fitP}.  
\end{equation}
For a lattice spacing $a$, this corresponds to the assumption that an
effective correlation length \be \xi(\beta)\,=\,a/\mu(\beta) \ee leads
to FSE which decay exponentially in the ratio $L/\xi$, where
$L\,=\,a\,\ns$ is the spatial extent of the lattice. By fitting
Eq. (\ref{fitP}) to the data, we determined $A(\beta)$, the plaquette
value on a spatially infinite lattice, the weight $B(\beta)$ of the
exponential component, and the effective mass $\mu (\beta)$ in lattice
units.  Depending on the type of plaquette considered, the quantities
$A,B,\mu$ are given the corresponding subscript $0$ or $\tau$ or $S$.

The general features of our findings on these FSE are the followings.
\begin{enumerate}
\item The parametrization (\ref{fitP}) is always adequate. As an
illustration, Fig.~(\ref{ps_beta}) shows how the data for $\av{P_S(\beta,\ns)}$
approach their $\ns = \infty$ limit, together with fits to Eq.  (\ref{fitP}).
\begin{figure}
\begin{center}
\includegraphics[width=12cm]{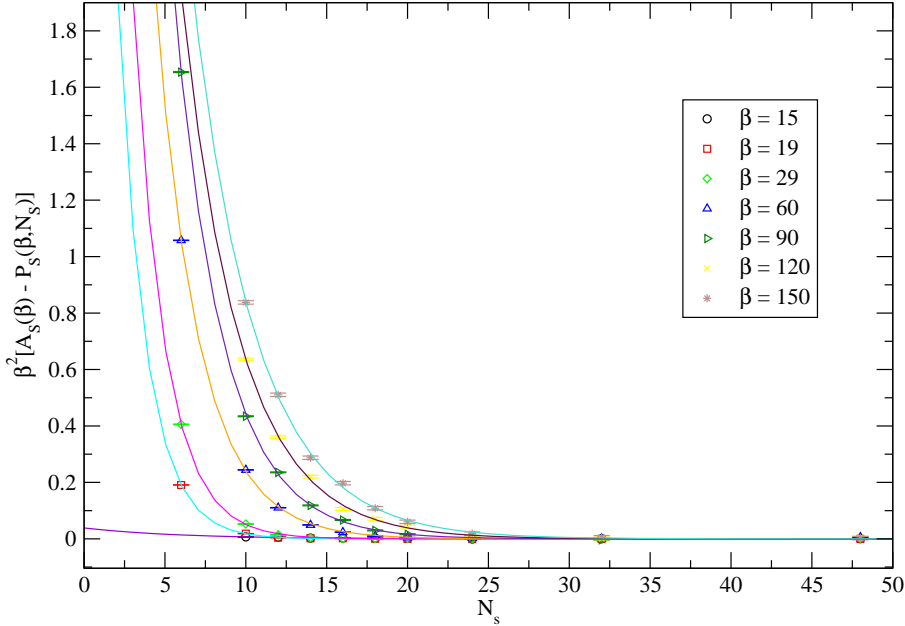}
\end{center}
\caption{\label{ps_beta} The average spatial plaquette
$\av{P_S(\beta,\ns)}$ approaches its limit $\av{P_S(\beta,\infty)}$
from below, exponentially in $\ns$. For clarity of the figure, the
quantity $A_S(\beta)\,-\,\av{P_S(\beta,\ns)}$ has been multiplied by
$\beta^2$.}
\end{figure}

\item Close to $\beta_c$, the $B$ coefficients are of opposite signs in
$\pt$ and $\ps$, and a partial cancellation of the FSE occurs. They
are of the same positive sign in $P_0$ and $\ps$, which again lowers the
overall FSE.

\item The mass parameter $\mu_\tau$, found to be around
.45 for $\beta/\nt$ large is poorly determined.
On the contrary, $\mu_S$ is well determined, and it lies in between
the two screening masses $M_S$ and $M_P$ found in \cite{bialas}, as 
illustrated in Fig.~(\ref{msSP}). In the range $\beta \in [15:34]$ for
which data on glueball masses exist \cite{teper}, we find that $\mu_0$ is 
lower than those, but only by about $20\%$. In the whole range 
$\beta \in [15:150]$ we have approximate scaling, i.e.  
$\beta\,\mu_0(\beta) \propto \mu_0(\beta)/a$ is nearly constant.

\item Finally the comparison with the statistical errors  of the 
effects due to $\ns$ finite shows that the latter can be neglected
in  our final measurements, all performed with the aspect ratio $\xi=\ns/\nt=8$. 
In the present exploration, we may compare the cases $\xi=4$ and $\xi=8$
corresponding respectively to $\ns$=16 and 32. Then for each $\av{P}$,
the $\ns-$dependent term in (\ref{fitP}) is reduced by a factor 
$\exp(16\,\mu)$ if $\xi=8$, that is at least by two orders of magnitude
in typical cases.  Moreover,as mentioned in point 2 above, the contributions
of these terms to $\nt^3\,\ds$ from different plaquettes tend to cancel 
each other. On the contrary the statistical errors are much larger in $\nt^3\,\ds$
than in $\av{P}$ in relative value.   

\begin{figure}
\begin{center}
\includegraphics[width=12cm]{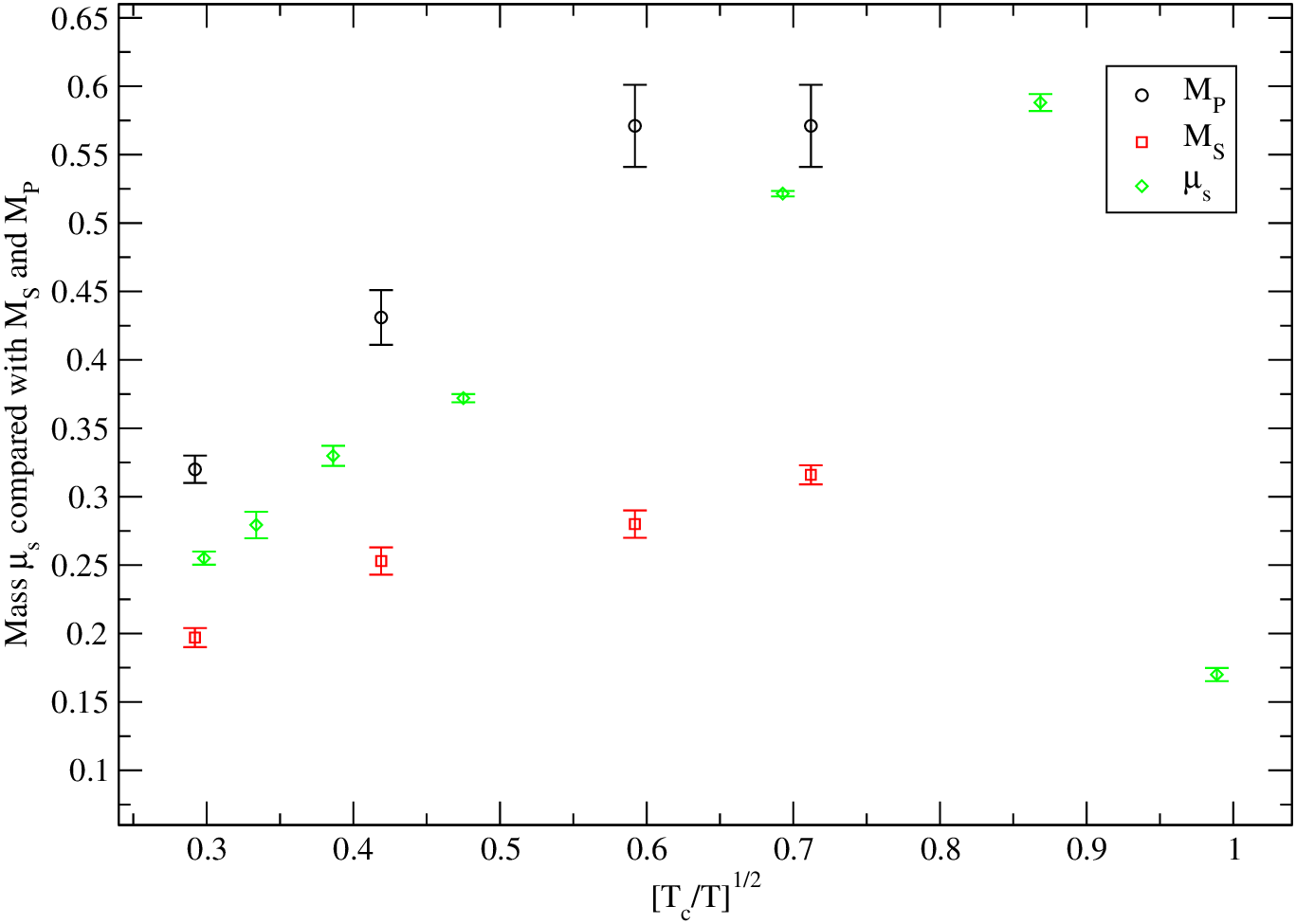}
\end{center}
\caption{\label{msSP}}
Comparison of the inverse correlation length $\mu_S$ appearing in the
FSE for $P_S$ with the screening masses $M_S$ and $M_P$ determined in
Ref.\cite{bialas}.
\end{figure}
\end{enumerate}

\medskip\noindent
The results for $\nt^3\,\ds$, for those lattice sizes of particular interest
in the
final computation of the thermodynamical variables ($\nt=4,6,8$) are presented in
figure~\ref{fig:ds}. Their numerical values are given in the tables of the appendix, where 
the case $\nt=2$ is also included.
\medskip\noindent
\medskip\noindent

\begin{figure}
\begin{center}
\includegraphics[height=14cm]{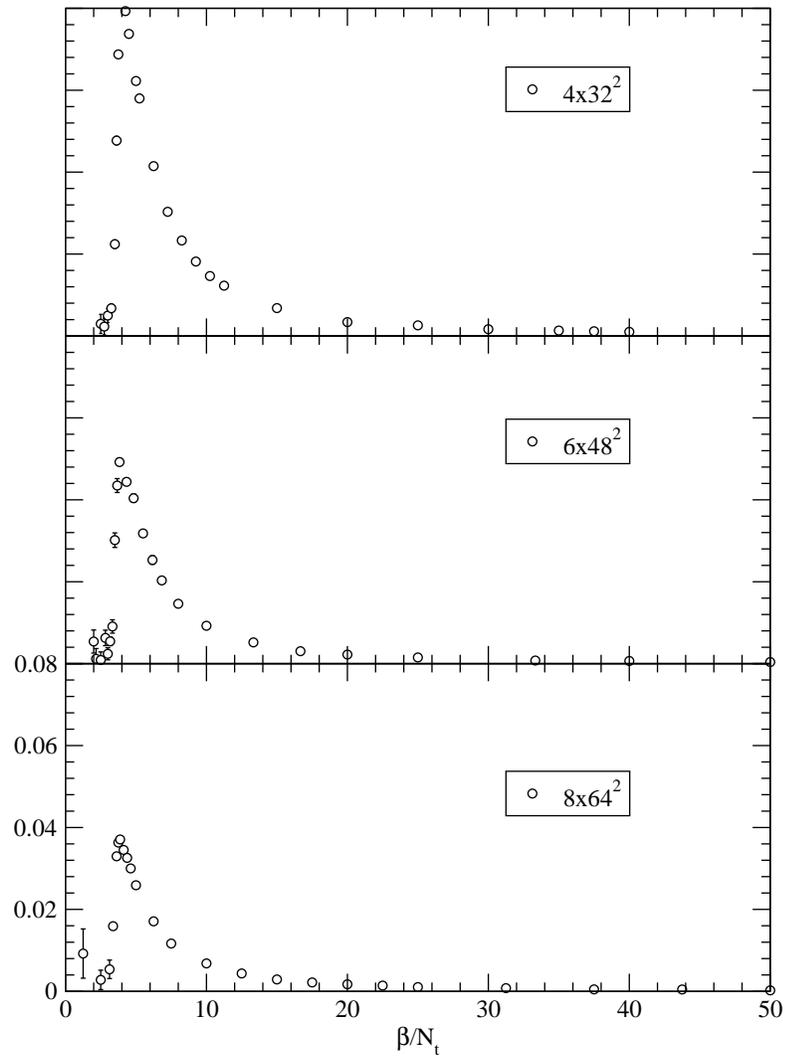}
\end{center}
\caption{\label{fig:ds}The average plaquette combination $\nt^3\Delta S$ for
$\nt=4,6,8$. Their values and errors (in most cases smaller than
 the point sizes) are tabulated in the appendix.
  On each plot the horizontal scale is $\beta\,/\,\nt$
 for easier comparison. The vertical scale is the
  same on all plots.}
\end{figure}

\section{Thermodynamics}

From the results presented, just above for $\nt^3\, \ds$ 
as a function of the lattice coupling $\beta$ and in section 3 
for the scaling function relating $\beta$ to the temperature, 
the three thermodynamical variables
$\epsilon,\,p$ and $s$ can be explicitly determined from the formulae of
section 2.

For definitness, we choose to measure the temperature in units of the
critical temperature. The temperature variable and 
the scaling function are then denoted
\bea
t\,&=&\,\frac{T}{T_c},  \\
\beta\,&=&\,\beta(t).
\eea

On a given $\nt\,\times\,\ns^2$ lattice, the dimensionless energy and 
pressure densities $\epsilon/T^3$ and $p/T^3$ 
follow from Eqs. (\ref{platt}, \ref{anomaly}) rewritten as
\bea \label{thermo}
\frac{\epsilon - 2 p}{T^3}(t) & = & 
  \nt^3 \,\ds\,\biggl(\beta (t) \biggr)\,t\frac{d\beta}{dt}, \\
\frac{p(t)}{T^3}\,&=&\,\nt^3\,\int^{\beta(t)}_{\beta(t_0)}
\, d \beta^\prime \ds(\beta'). 
\eea
In this integral, we choose $t_0\,=\,0.6$, a temperature below
which the integrand is always negligible. From these equations
one directly obtains $p/T^3$ and $\epsilon/T^3$;  the entropy density
$s/T^2$ follows from  (\ref{entropy}).

In order to use Eqs. (\ref{thermo}) in practice, and 
in particular to
perform the integral defining the pressure conveniently, we have 
interpolated $\nt^3\,\ds (\beta(t))$ via a smooth function $f(t)$.
In the absence of any theoretical ansatz for the shape observed in 
figure~\ref{fig:ds}, 
we choose the simplest parametrization adapted to the description of
a sharp rise in the vicinity of the critical point $t$=1, and the  
$t^{-2}$  fall off at large t 
expected from Eqs.(\ref{beta3},\ref{fastb}). 
The following function constitutes an accurate  
representation of $\nt^3\,\ds (\beta(t))$ within errors 
\begin{equation} 
f(t)=\frac{a_1 t^2}{1+a_2 t^4}\frac{1+a_3\, e^{a_5(t-1)}}
{1+a_4\, e^{a_5(t-1)}}. \label{ffit}
\end{equation}

\begin{figure}
\begin{center}
\includegraphics[width=12cm]{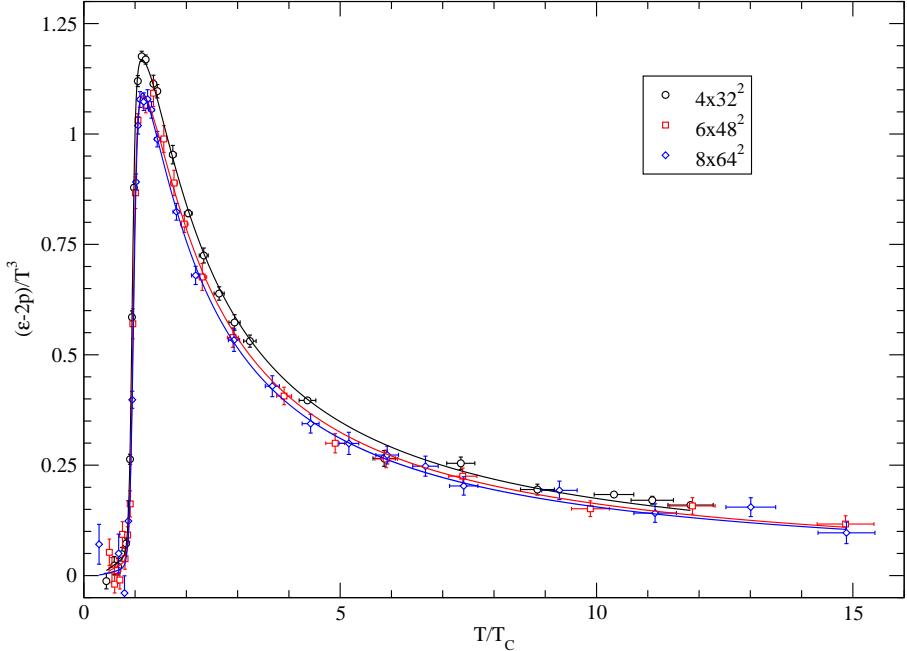}
\end{center}
\caption{\label{fig:em2p_fit.eps}Trace of the energy momentum
tensor.}
\end{figure}

The adequacy of this function for our purposes is illustrated by the
curves drawn through the data points in 
figure~\ref{fig:em2p_fit.eps}. The horizontal error bars 
represent the errors in the change of variable $\beta\,\to\,t=T/T_c$
described in section 3. They are of the order of 3 percents  of $t$. 

One may worry that choosing a specific function like (\ref{ffit}) introduces 
a systematic error into subsequent calculations.  
A different choice was tried during this investigation. To
the extent that it provided a comparable agreement with the data to which it
was fitted, it leads, for example for the pressure, to differences much
smaller than the statistical errors. However one must keep in mind that
although $f(t)$ constitutes a good {\it {interpolation}} of the data 
within errors, the systematic errors attached to its use above 
$t\approx 15$ are not controlled. 

\begin{figure}
\begin{center}
\includegraphics[width=12cm]{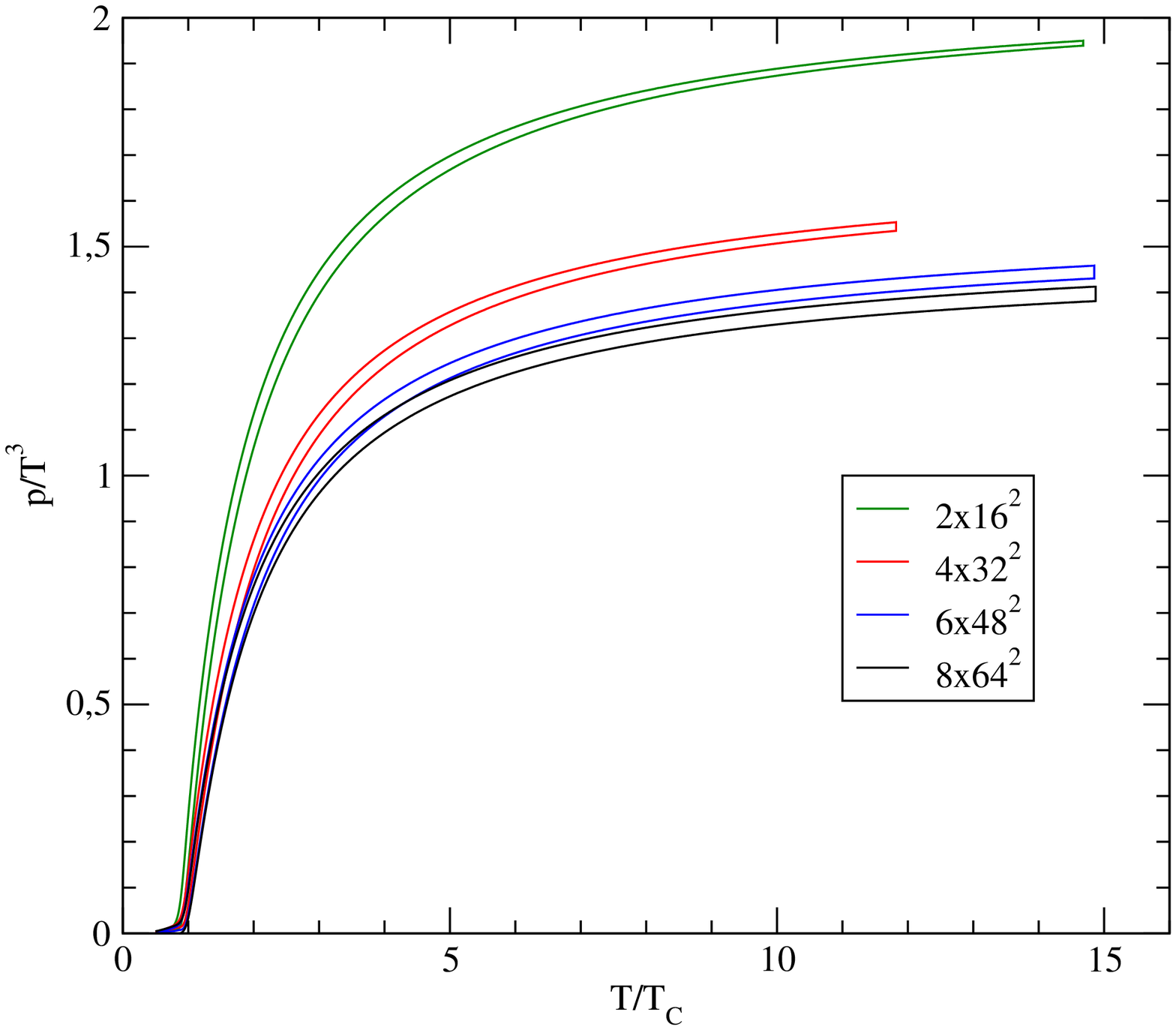}
\end{center}
\caption{\label{fig:press}The pressure measured on lattices with 
$\nt\,=\,2,4,6,8$ and $\ns\,=\,8\,\nt$}
\end{figure}

\begin{figure}
\begin{center}
\includegraphics[width=12cm]{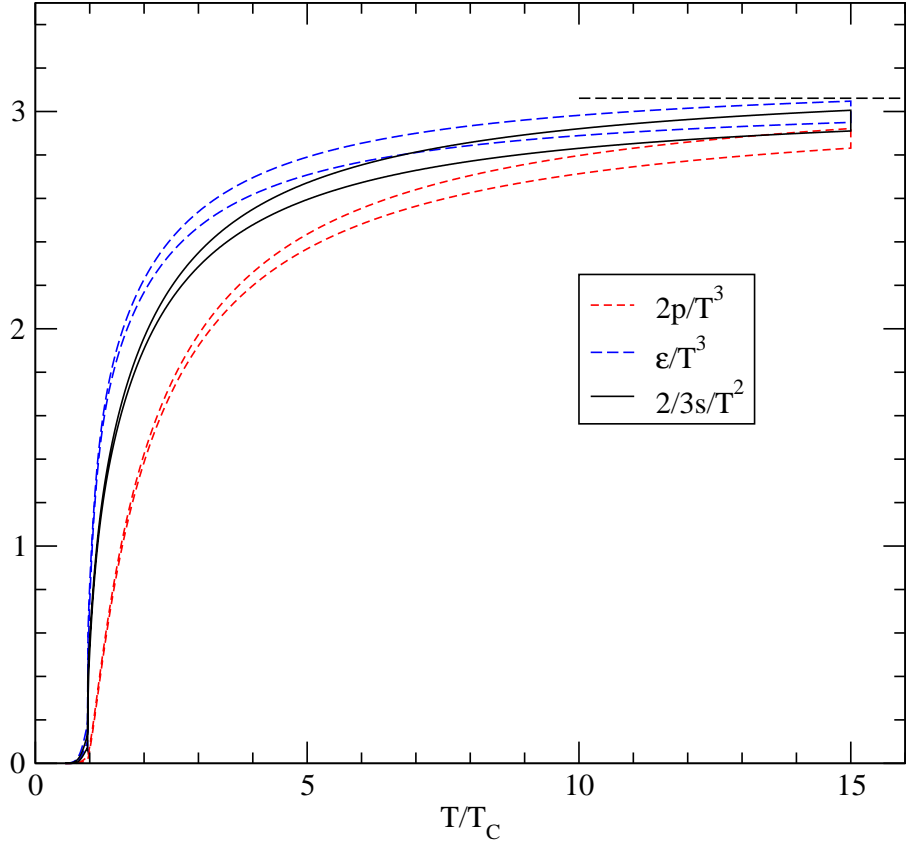}
\end{center}
\caption{\label{fig:cont}The pressure, and the energy and entropy densities 
for the largest lattice size ($8\text{x}64^2$).}
\end{figure}

Our final results for the pressure at $\nt\,=\,2,4,6,8$ with $\ns\,=\,8\,\nt$
are shown in figure~\ref{fig:press}. The errors on the scale $t$, which we
showed in figure~\ref{fig:em2p_fit.eps}, do not affect the pressure,
which varies slowly where they become important (at large $t$). For the largest lattice considered
($\nt\,=\,8$), figure~\ref{fig:cont} compares the three thermodynamical 
variables,  after multiplication by coefficients such  that they 
would be the same in the free theory (see Eq.
\ref{freep}). The continuum Stefan-Boltzmann value for a free gas 
(horizontal dotted line) is also shown.

All the errors on the quantities plotted in figures~\ref{fig:em2p_fit.eps},
\ref{fig:press} and \ref{fig:cont} have been
estimated using the bootstrap method, similarly to the case described 
in section 3 for the calculation of $T/\sqs$.

\medskip\noindent
\medskip\noindent
\medskip\noindent

Let us now comment these results, whose extensions towards the
continuum limit $1/\nt\,=\,0$ and to larger temperatures are highly
desirable.
A look at the results suggests that the
continuum limit is nearly reached at $\nt = 8$, the more so $t$ is large.
We observe that the values of  
$(\epsilon-2p)/T^3$, at least above $t \approx 5$ hardly distinguish between
$\nt\,=\,6$ and 8, and that the difference between the corresponding pressures
does not exceed $\approx 2\sigma$. More quantitative
statements however are hard to produce.
Attempts at fitting  quadratic functions in
$1/\nt$ or $1/\nt^2$ to 
the pressure at fixed $t$ remained inconclusive, with no strong
constraint on the continuum limit.

We then turn to the question of the high temperature behaviour. We
already noticed (\ref{beta3}) that the tail at large $\beta$ of 
$\nt^3\,\ds$ was compatible with $1/\beta^2$, 
which implies via (\ref{fastb})  
\be
\frac{\epsilon-2p}{T^3} \approx \frac{b}{t}; \quad t>t_1. \label{hight}
\ee
for some large enough $t_1$. Since this appears to be true
for any $\nt$, it is reasonable to expect that it is also valid in the
continuum. We thus formulate the following ansatz: {\it {For $t$ larger than
some $t_{min}$, the continuum limit of $(\epsilon-2p)/T^3$ is close to its
value measured at $\nt\,=\,8$  and it can be parametrized there by $b/t$. }}
We then obtain from Eqs. (\ref{thermo}, \ref{fastb})

\begin{table}
\begin{center}
\begin{tabular}{||c||c|c|c||c|c||}
\hline\hline
$t_{min}$ & $b$ & $p/T^3(t_{min})$& $p/T^3(\infty)$\\
\hline\hline
4.0       & 1.81(7) & 1.13(2)&1.58\\
6.0       & 1.7(1)& 1.27(2)&1.56\\
8.0       & 1.6(2)& 1.34(2)&1.53\\
\hline\hline
\end{tabular}
\caption{\label{tab:a}}
\end{center}
\end{table}

\bea
\frac{p}{T^3}(t)&=&
\int_{t_0}^{t}\text{d}t \frac{1}{t}\frac{\epsilon-2p}{T^3}(t), \label{p}\\
&=&a\,-\,\frac{b}{t}, \label{pt}\\
a&=&\frac{p}{T^3}(t_{min})\,+\,\frac{b}{t_{min}}. \label{a} 
\eea
For $\nt=8$, the data  between $t\approx 4$
and $t=15$ (the largest value explored) is compatible with a behaviour of
the type (\ref{hight}). Suppose now that this remains true above $t$=15. 
 Then from (\ref{pt},\ref{a}), we have
\be
\frac{p}{T^3}(t=\infty) \,=\, \frac{p}{T^3}(t_{min})\,+\,\frac{b}{t_{min}}.
\ee
Given $t_{min}$, we get $b$ from $(\epsilon-2p)/T^3(t_{min})$, and
$p/T^3(t_{min})$ is known  from (\ref{p}), so that a
consistency check is that $p/T^3(t=\infty)$ is independent of 
$t_{min}$. In table ~\ref{tab:a}, we report the values found for $b$
and $p/T^3(t=\infty)$ for $t_{min}$ = 4,6 and 8. We did not
try to estimate any error on $p/T^3(t=\infty)$, certainly affected by
uncontrolled systematics. Its value is stable with respect to $t_{min}$, and
although it is obtained for $\nt$ finite,  
it is found close to the continuum Stefan-Boltzmann  value for a free gas, 1.53.

\section{Conclusions}

In this article we have determined the thermodynamical quantities for SU(3) gauge theory
in two spatial dimensions. This determination has been performed through lattice simulations, 
measuring the thermodynamical variables as functions of the lattice coupling $\beta$.
We then constructed the scaling function connecting $\beta$
to the temperature expressed in units of various physical quantities, the square root $\sqs$ of the zero temperature
 string tension, the coupling constant $g^2$ of the continuum theory and the critical temperature $T_c$. In particular,
 we determined the ratios $T_c/\sqs$ and $T_c/g^2$ to be respectively 1.00(4) and 0.55(2).
 This construction has
 been discussed in detail, using published data on the string tension and on the critical lattice coupling
 $\beta_c$ .We finally kept $T_c$ as the temperature scale and presented the thermodynamical variables as functions of 
$T/T_c$. 

We have determined the finite volume $\ns$ effects at fixed bare coupling constant $\beta$ by a careful
analysis and shown that they are negligible compared to the statistical errors. We have further shown that
those finite size effects fall off exponentially with $\ns$, and that the coefficient in the exponent
multiplying $\ns$ is consistent within about $20\%$ with the longest correlation lengths expected on the various lattices,
namely the inverse of the lowest glue ball mass on the zero temperature lattices,
and the lowest screening length on the finite temperature lattices. This has lead us to choose the aspect ratio
$\ns/\nt =8$, for which the finite size effects are much smaller then the statistical errors.

We investigated the finite lattice spacing effects by simulating lattices with extension in the
temperature direction $\nt =2,4,6,8$. We found that $\nt=8$ is large enough to be near to the continuum
limit. We have not presented attempts to extrapolate the thermodynamical variables to the continuum, because additional
systematic errors appear. Instead
 we have analyzed the thermodynamical variables for $\nt=8$. We find that for large enough temperatures, well above
the phase transition region, the 
pressure can be represented by  a
very simple function, namely $p=T^3\,(a - b T_c/T)$. Using the thermodynamical identities for a large homogeneous system we obtain
the corresponding formulae for the other thermodynamical quantities.

Improving our findings in view of a quantitative determination of the continuum limit requires a more precise 
analysis of the $\nt$ dependence of the basic quantity $\nt^3\,\ds$. Larger statistics and/or
larger lattices are possible ways in this direction.

\section{Acknowledgements}

This work is supported through by the European Community's
Sixth Framework Programme, network contracts MRTN-CT-2004-005616(ENRAGE) and
CT-2004-517186(COCOS) and KBN grant 1P03B-04029. 
We thank the IPhT (CEA-Saclay) and the Institute of Physics at the Jagiellonian
University ( Krakow) for their
kind hospitality. Furthermore B.P. is grateful for the hospitality of the TIFR, Mumbai 
and for discussions with R. Gavai and Sourendu Gupta. P.B. thanks Fakult\"at f\"ur a
Physik (University of Bielefeld)
for kind hospitality and for discussions with O. Kaczmarek and E. Laermann.

\appendix

\section{Data}
\label{sec:data}

The data were produced from a standard $SU(3)$ MC program. We 
used a mixture of heatbath  and overrelaxation
 moves,  with typically one heatbath sweep after
2 to 4 overrelaxation sweeps. All the simulations were done on the
COCOS opteron cluster in the Institute of Physics, Jagiellonian
University.  This appendix presents the data for the plaquette
averages \mbox{$\av{P_x}=1-\frac{1}{3}\tr\av{U_{P_x}}$}. The subscripts
$S$ and $\tau$ denote spacelike and timelike plaquettes respectively, on
finite temperature lattices ($\nt<\ns$). The subscript zero denotes 
plaquettes on zero temperature lattices ($\nt=\ns$). For each of
the $\nt\,\times\,\ns^2$ or $\ns^3$ lattices, as functions of $\beta$
we list the values of $\av{\pt}$, 
$\av{\ps}$, $\av{\po}$, and $\av{\nt ^3 \ds}$ from which
the thermodynamical variables are finally computed. The errors 
were estimated using standard methods. 
We took care of strong correlations between $\pt$ and $\ps$, 
by first measuring the statistical errors for the combination $2\pt\,+\,\ps$,
and adding them quadratically to those of the independent quantity 3$\po$
to finally get the errors on $\av{\ds}\,=\,\av{3\po-(2\pt\,+\,\ps)}$

\begin{table}[H]
\begin{center}
\begin{tabular}{||D{.}{.}{3}||D{.}{.}{13}|D{.}{.}{13}|D{.}{.}{13}|D{.}{.}{13}||}
\hline\hline
\multicolumn{1}{||c||}{$\beta$} & \multicolumn{1}{c|}{$\av{\pt}$} & \multicolumn{1}{c|}{$\av{\ps}$} & \multicolumn{1}{c|}{$\av{\po}$} & \multicolumn{1}{c||}{$\av{\nt^3\,\ds}$} \\\hline\hline
\input {plaq.2x16fin2}
\hline\hline
\end{tabular}
\end{center}
\caption{\label{data:2x16}Plaquette data for lattices of size $2\text{x}16^2$ and $16^3$.}
\end{table}

\begin{table}[H]
\begin{center}
\begin{tabular}{||D{.}{.}{3}||D{.}{.}{13}|D{.}{.}{13}|D{.}{.}{13}|D{.}{.}{13}||}
\hline\hline
\multicolumn{1}{||c||}{$\beta$} & \multicolumn{1}{c|}{$\av{\pt}$} & \multicolumn{1}{c|}{$\av{\ps}$} & \multicolumn{1}{c|}{$\av{\po}$} & \multicolumn{1}{c||}{$\av{\nt^3\,\ds}$} \\\hline\hline
\input {plaq.4x32fin2}
\hline\hline
\end{tabular}
\end{center}
\caption{\label{data:4x32}Plaquette data for lattices of size $4\text{x}32^2$ and $32^3$.}
\end{table}

\begin{table}[H]
\begin{center}
\begin{tabular}{||D{.}{.}{3}||D{.}{.}{13}|D{.}{.}{13}|D{.}{.}{13}|D{.}{.}{13}||}
\hline\hline
\multicolumn{1}{||c||}{$\beta$} & \multicolumn{1}{c|}{$\av{\pt}$} & \multicolumn{1}{c|}{$\av{\ps}$} & \multicolumn{1}{c|}{$\av{\po}$} & \multicolumn{1}{c||}{$\av{\nt^3\,\ds}$} \\\hline\hline
\input {plaq.6x48fin2}
\hline\hline
\end{tabular}
\end{center}
\caption{\label{data:6x48}Plaquette data for lattices of size $6\text{x}48^2$ and $48^3$.}
\end{table}

\begin{table}[H]
\begin{center}
\begin{tabular}{||D{.}{.}{3}||D{.}{.}{13}|D{.}{.}{13}|D{.}{.}{13}|D{.}{.}{13}||}
\hline\hline
\multicolumn{1}{||c||}{$\beta$} & \multicolumn{1}{c|}{$\av{\pt}$} & \multicolumn{1}{c|}{$\av{\ps}$} & \multicolumn{1}{c|}{$\av{\po}$} & \multicolumn{1}{c||}{$\av{\nt^3\,\ds}$} \\\hline\hline
\input {plaq.8x64fin2}
\hline\hline
\end{tabular}
\end{center}
\caption{\label{data:8x64}Plaquette data for lattices of size $8\text{x}64^2$ and $64^3$.}
\end{table}


\end{document}

%% file: sigma.tex
\begin{tabular}{||l|l|l|l|l|l||}
\hline\hline
$\beta$ & $N_s$ &$F_\sigma(\beta)=a\sqrt{\sigma} $  & ref.\\
\hline\hline
8.156   & 16    & 0.5677(19)    &\cite{lego}  \\
10      & 24    & 0.42443(70)   &\cite{lego}  \\
12      & 24    & 0.33679(21)   &\cite{lego}  \\
14      & 24    & 0.27885(13)   &\cite{lego}  \\
14.7172	&       & 0.26101(9)    &\cite{teper2}\\
15	& 24    & 0.2570(15)    &\cite{teper}\\
18	& 32    & 0.20712(24)   &\cite{lego}\\
19	& 32    & 0.19447(23)   &\cite{lego}\\
21	&       & 0.173948(75)  &\cite{teper2}\\
22	& 32    & 0.16555(24)   &\cite{lego}\\
24	& 32    & 0.15112(23)   &\cite{lego}\\
28	& 32    & 0.1275(2)     &\cite{teper}\\
34	& 40    & 0.10379(26)   &\cite{teper}\\
40	&       & 0.087046(75)  &\cite{teper2}\\
50	& 48    & 0.07021(5)    &\cite{lego}\\
\hline\hline
\end{tabular}

%% file: plaq.2x16fin2.tex
   4.00  &       0.715487(18) &       0.715548(24) &       0.715527(11) &       0.00048(45) \\
   6.00  &       0.541947(21) &       0.542644(25) &       0.542474(14) &       0.00709(57) \\
   6.50  &       0.498643(22) &       0.499733(24) &       0.499415(14) &       0.00981(58) \\
   7.00  &       0.457750(21) &       0.459285(23) &       0.458821(14) &       0.01342(57) \\
   7.50  &       0.420260(21) &       0.422586(22) &       0.422054(13) &       0.02446(55) \\
   8.00  &       0.382967(42) &       0.389985(20) &      0.3896419(97) &       0.10405(77) \\
   8.50  &       0.346390(21) &       0.362726(18) &       0.361541(11) &       0.23293(49) \\
   9.00  &       0.321171(16) &       0.340272(16) &      0.3372283(98) &       0.23257(41) \\
   9.50  &       0.300423(14) &       0.320794(16) &      0.3160913(90) &       0.21306(36) \\
  10.00  &       0.282596(13) &       0.303523(15) &      0.2975888(48) &       0.19242(29) \\
  11.00  &       0.253093(11) &       0.274289(13) &      0.2666359(42) &      0.15545(25) \\
  13.00  &      0.2100280(87) &       0.230317(11) &      0.2211848(33) &      0.10545(20) \\
  14.00  &      0.1936981(79) &       0.213253(10) &      0.2039362(31) &      0.08928(19) \\
  15.00  &      0.1797770(73) &      0.1986216(94) &      0.1892295(28) &      0.07610(17) \\
  16.00  &      0.1677617(67) &      0.1858851(87) &      0.1765258(26) &      0.06535(16) \\
  17.00  &      0.1572549(63) &      0.1746807(82) &      0.1654467(24) &      0.05720(15) \\
  18.00  &      0.1479903(59) &      0.1647495(78) &      0.1556880(22) &      0.05067(14) \\
  20.00  &      0.1324407(53) &      0.1479638(70) &      0.1392945(20) &      0.04031(12) \\
  21.00  &      0.1258401(49) &      0.1407836(66) &      0.1323299(19) &      0.03621(12) \\
  25.00  &      0.1049274(41) &      0.1179602(56) &      0.1103147(16) &      0.025033(96) \\
  29.00  &      0.0900011(35) &      0.1015174(48) &      0.0945963(13) &      0.018155(82) \\
  33.00  &      0.0788007(31) &      0.0891024(42) &      0.0828101(12) &      0.013813(72) \\
  37.00  &      0.0700760(27) &      0.0793988(38) &      0.0736427(10) &      0.011019(64) \\
  41.00  &      0.0631038(24) &      0.0716004(34) &     0.06630419(53) &      0.008836(54) \\
  45.00  &      0.0573848(22) &      0.0652062(31) &     0.06029866(84) &      0.007362(52) \\
  60.00  &      0.0428417(16) &      0.0488355(23) &     0.04501735(62) &      0.004265(39) \\
  80.00  &      0.0320233(12) &      0.0365998(17) &     0.03364921(47) &      0.002410(29) \\
 100.00  &     0.02556863(98) &      0.0292673(14) &     0.02686572(37) &     0.001541(23) \\
 120.00  &     0.02127680(81) &      0.0243810(12) &     0.02235950(44) &     0.001152(21) \\
 200.00  &     0.01273399(49) &     0.01462132(70) &     0.01338190(26) &     0.000451(12) \\

%% file: plaq.4x32fin2.tex
8.000 & 0.389671(18) & 0.389672(21) & 0.389649(12) & -0.0022(37) \\ 
10.000 & 0.297569(13) & 0.297594(15) & 0.2975976(83) & 0.0030(25) \\ 
11.000 & 0.266621(11) & 0.266656(13) & 0.2666393(73) & 0.0023(22) \\ 
12.000 & 0.2416874(71) & 0.2417314(82) & 0.2417244(64) & 0.0050(16) \\ 
13.000 & 0.2211207(52) & 0.2212044(61) & 0.2211845(41) & 0.0068(11) \\ 
14.000 & 0.2037477(50) & 0.2039560(56) & 0.2039335(37) & 0.0224(10) \\ 
15.000 & 0.1886596(30) & 0.1892923(33) & 0.1892303(13) & 0.06874(51) \\ 
16.000 & 0.1758268(40) & 0.1766757(48) & 0.1765290(32) & 0.08113(84) \\ 
17.000 & 0.1647251(38) & 0.1656690(45) & 0.1654499(29) & 0.07931(78) \\ 
18.000 & 0.1549712(35) & 0.1559739(42) & 0.1556893(27) & 0.07374(72) \\ 
20.000 & 0.1386185(53) & 0.1396605(66) & 0.1392895(34) & 0.0622(10) \\ 
21.000 & 0.1316761(50) & 0.1327346(62) & 0.1323285(31) & 0.05800(95) \\ 
25.000 & 0.1097556(41) & 0.1107875(52) & 0.1103143(27) & 0.04147(80) \\ 
29.000 & 0.0941234(14) & 0.0950946(18) & 0.09460468(62) & 0.03032(25) \\ 
33.000 & 0.0823971(31) & 0.0833061(39) & 0.0828209(20) & 0.02331(60) \\ 
37.000 & 0.0732736(27) & 0.0741220(34) & 0.0736508(18) & 0.01817(52) \\ 
41.000 & 0.0659708(24) & 0.0667588(31) & 0.0663092(16) & 0.01463(47) \\ 
45.000 & 0.0599962(20) & 0.0607343(24) & 0.06030640(65) & 0.01229(34) \\ 
60.000 & 0.04478745(25) & 0.04538028(32) & 0.04502049(34) & 0.006806(72) \\ 
80.000 & 0.0334775(12) & 0.0339454(16) & 0.03365195(77) & 0.00338(23) \\ 
100.000 & 0.02672828(77) & 0.02711023(97) & 0.02686910(35) & 0.00258(14) \\ 
120.000 & 0.02224522(58) & 0.02256883(75) & 0.02236161(25) & 0.00164(10) \\ 
140.000 & 0.01904894(21) & 0.01933046(27) & 0.01914982(17) & 0.001326(45) \\ 
150.000 & 0.01777228(36) & 0.01803633(46) & 0.01786617(16) & 0.001149(62) \\ 
160.000 & 0.01665607(19) & 0.01690470(24) & 0.01674434(15) & 0.001008(39) \\ 

%% file: plaq.6x48fin2.tex
12.000 & 0.2417134(38) & 0.2417175(45) & 0.2417226(35) & 0.0054(29) \\ 
13.000 & 0.2211807(34) & 0.2211791(40) & 0.2211817(31) & 0.0012(26) \\ 
14.000 & 0.2039355(29) & 0.2039378(34) & 0.2039331(19) & -0.0016(19) \\ 
15.000 & 0.1892252(28) & 0.1892281(35) & 0.1892265(26) & 0.0009(22) \\ 
16.000 & 0.1765245(24) & 0.1765294(30) & 0.1765261(16) & -0.0007(16) \\ 
17.000 & 0.1654377(25) & 0.1654423(30) & 0.1654478(23) & 0.0063(19) \\ 
18.000 & 0.1556821(23) & 0.1556927(28) & 0.1556901(21) & 0.0024(17) \\ 
19.000 & 0.1470158(17) & 0.1470316(20) & 0.1470297(11) & 0.0055(11) \\ 
20.000 & 0.1392652(26) & 0.1392938(32) & 0.1392882(14) & 0.0091(16) \\ 
21.000 & 0.1322566(28) & 0.1323336(32) & 0.1323301(16) & 0.0301(18) \\ 
22.000 & 0.1259308(26) & 0.1260530(31) & 0.1260376(16) & 0.0435(17) \\ 
23.000 & 0.1201921(16) & 0.1203390(18) & 0.12031923(83) & 0.04922(79) \\ 
26.000 & 0.1057875(20) & 0.1059611(23) & 0.10591516(74) & 0.04438(79) \\ 
29.000 & 0.0944815(17) & 0.0946652(20) & 0.09460286(65) & 0.0404(10) \\ 
33.000 & 0.0827075(15) & 0.0828960(18) & 0.08281948(71) & 0.03179(92) \\ 
37.000 & 0.0735500(13) & 0.0737324(16) & 0.07365035(80) & 0.02530(86) \\ 
41.000 & 0.06622256(75) & 0.06639958(91) & 0.06631278(43) & 0.02032(47) \\ 
48.000 & 0.0563954(11) & 0.0565549(13) & 0.05647074(47) & 0.01462(63) \\ 
60.000 & 0.04495852(65) & 0.04510118(80) & 0.04502038(34) & 0.00925(39) \\ 
80.000 & 0.03360653(36) & 0.03372051(45) & 0.03365228(25) & 0.00519(24) \\ 
100.000 & 0.02683241(36) & 0.02692816(44) & 0.02686872(16) & 0.00304(21) \\ 
120.000 & 0.02233119(22) & 0.02241268(28) & 0.02236182(17) & 0.00223(15) \\ 
150.000 & 0.01784145(19) & 0.01790881(24) & 0.01786618(13) & 0.00152(13) \\ 
200.000 & 0.01336444(13) & 0.01341620(17) & 0.01338281(10) & 0.000763(91) \\ 
240.000 & 0.01112982(12) & 0.01117361(15) & 0.011145438(75) & 0.000661(77) \\ 
300.000 & 0.00889827(10) & 0.00893393(13) & 0.008910796(50) & 0.000391(62) \\ 

%% file: plaq.8x64fin2.tex
10.000 & 0.2975884(59) & 0.2976005(72) & 0.2975867(40) & 0.0092(60) \\ 
20.000 & 0.13928938(92) & 0.1392889(12) & 0.13929140(79) & 0.0028(25) \\ 
23.000 & 0.12031707(79) & 0.1203197(10) & 0.12031754(73) & -0.0019(21) \\ 
25.000 & 0.11031138(86) & 0.1103157(11) & 0.11031518(72) & 0.0054(19) \\ 
27.000 & 0.10184029(88) & 0.1018568(11) & 0.10185383(82) & 0.01591(71) \\ 
29.000 & 0.09456979(81) & 0.0946071(11) & 0.09460236(76) & 0.03296(65) \\ 
30.000 & 0.09131473(79) & 0.0913588(10) & 0.09135313(72) & 0.03632(72) \\ 
31.000 & 0.08828064(88) & 0.0883287(11) & 0.08831814(80) & 0.03708(61) \\ 
33.000 & 0.08277950(54) & 0.08283196(70) & 0.08281902(50) & 0.03451(62) \\ 
35.000 & 0.07792616(50) & 0.07798094(66) & 0.07796591(47) & 0.03257(58) \\ 
37.000 & 0.07361228(63) & 0.07366982(82) & 0.07365076(58) & 0.03001(55) \\ 
40.000 & 0.06796965(58) & 0.06802736(76) & 0.06800643(53) & 0.02589(46) \\ 
50.000 & 0.05414363(41) & 0.05419933(54) & 0.05417403(37) & 0.01708(44) \\ 
60.000 & 0.04499613(38) & 0.04504689(50) & 0.04502046(35) & 0.01166(32) \\ 
80.000 & 0.03363355(33) & 0.03367621(43) & 0.03365276(30) & 0.00680(32) \\ 
100.000 & 0.02685397(17) & 0.02689045(22) & 0.02686885(15) & 0.00436(23) \\ 
120.000 & 0.02234902(13) & 0.02238078(17) & 0.02236201(12) & 0.00290(18) \\ 
140.000 & 0.01913902(16) & 0.01916675(21) & 0.01914968(15) & 0.00216(16) \\ 
160.000 & 0.01673517(14) & 0.01676024(19) & 0.01674472(13) & 0.00172(14) \\ 
180.000 & 0.01486781(12) & 0.01489035(16) & 0.01487629(11) & 0.00139(12) \\ 
200.000 & 0.01337535(11) & 0.01339589(15) & 0.01338306(10) & 0.00102(11) \\ 
250.000 & 0.010692216(90) & 0.01070908(12) & 0.010698237(84) & 0.000777(90) \\ 
300.000 & 0.008905754(75) & 0.008920037(99) & 0.008910804(69) & 0.000474(76) \\ 
350.000 & 0.007630792(65) & 0.007643064(84) & 0.007635106(60) & 0.000445(65) \\ 
400.000 & 0.006675004(80) & 0.00668602(10) & 0.006678986(93) & 0.000243(59) \\ 